\documentstyle[twocolumn,aps]{revtex}

\begin{document}

\draft


\title{Stability of  degenerate Cauchy horizons in black hole spacetimes}

\author{Rong-Gen Cai}
\address{Center for Theoretical Physics, Seoul National University,
        Seoul 151-742, Korea}

\maketitle

\begin{abstract}
In the multihorizon black hole spacetimes, it is possible  that there are  
degenerate Cauchy horizons  with vanishing surface gravities. 
We investigate the stability of the degenerate Cauchy horizon in black hole 
spacetimes. Despite the asymptotic behavior of spacetimes (flat, anti-de 
Sitter, or de Sitter), we find that the Cauchy horizon is stable against 
the classical perturbations, but unstable quantum mechanically.

\end{abstract}

The stability of Cauchy horizon (CH) in black hole spacetimes has been a
 subject of great interest during the last few  years. 
This issue is closely relevant  to the 
Penrose's cosmic censorship \cite{Pen1}, which states \cite{BMM}: the
evolution of generic initial conditions will always produce a globally
hyperbolic spacetime. The CH is the boundary of the Cauchy 
development of a maximal Cauchy surface. The inner horizon of the 
Reissner-Nordstr\"om (RN) black holes is a familiar example of CH's 
in black hole spacetimes.

Beyond the CH the predictability of general relativity is lost. Therefore,
the strong cosmic censorship will be violated in the RN black hole 
spacetimes if the CH is stable. Indeed the CH of the RN black holes has
 been proved to be unstable in the linear perturbations \cite{Chand}, 
nonlinear perturbations \cite{Brady1}, and 
exact backreaction calculations \cite{Hiscock}, and a scalar curvature 
singularity is formed there. The instability of the CH evades the 
loss-of-predictability problem and the scalar curvature singularity prevents
the further evolution of the spacetime beyond the CH. In this sense, the 
strong cosmic censorship still holds. The CH of the Kerr black holes is 
also unstable \cite{Kerr}.

The ``stable'' CH emerges  when the RN black holes are embedded in 
 de Sitter space. In this case, except for the CH and the event horizon 
of the black holes, a cosmological horizon may appear. 
When $\kappa _i \le \kappa _c$, the CH is proven to be classically 
 stable \cite{RND}, where $\kappa _i$ and $\kappa _c$ 
represent the surface gravities of the CH and the cosmological horizon,
respectively. The analysis based on the linear perturbation gives similar
property for the CH in the Kerr-de Sitter black holes \cite{Moss}. 
Using a two-dimensional RN-de Sitter model, Markovi\'c and Poisson 
\cite{Mark} have argued that the classical ``stable'' CH is still
 quantum-mechanically unstable, except for the set of measure zero 
spacetimes for which $\kappa_i=\kappa_c $, by calculating the renormalized 
expectation value of the stress-energy tensor associated with a 
conformal scalar field propagating in the two-dimensional background.

More recently, Brady, Moss, and Myers \cite{BMM} have made great progress
in the understanding of classical stability of the CH for the RN-de Sitter
spacetimes. They have noticed that, except for the well-known 
 generic radiation tail falling off exponentially, 
 there is an additional  influx resulting from outgoing modes 
 due to the backscattering off the spacetime curvature.
Although the latter is in the subleading order, it  produces a divergent
influx at the CH so that the CH becomes  unstable. Thus, 
the so-called counter-example of the strong cosmic censorship is 
completely disproved. Consequently, an interesting question arises 
whether a stable CH exists in the black hole spacetimes as a
 counter-example of the strong cosmic censorship, at least 
classically. In this paper  we present 
such an example  in which  the black hole has a  degenerate CH.

Let's begin our discussion by introducing two exact black hole solutions
possessing multihorizon structures. The first one is a two-dimensional
dilaton black hole, which comes from the following action
\begin{eqnarray}
\label{action1}
S &=& \int d^2x \sqrt{-g}\  e^{-2\phi} \left [ R+2 (\nabla \phi)^2 
      -\frac{1}{4}F_{\mu\nu}F^{\mu\nu} 
   \right.   \nonumber \\
   &+& \left. 2\ e^{2\phi}
    +\sum ^{k}_{n=2}a_ne^{2n\phi}\right ],
\end{eqnarray}
where $R$ denotes the curvature scalar, $F_{\mu\nu}$ is the Maxwell 
field and $a_n$ are constants. The dilaton potential may arise in the slop
expansion of some string theories.  The black hole solution is \cite{Cai}
\begin{eqnarray}
\label{solution1}
&& ds^2 =-f(r) ~dt^2 +f^{-1}(r)~dr^2, \\
&& \phi (r)= -\ln ~ r, \nonumber \\
&& F_{tr}=q/r^2, \nonumber \\
&& f(r)= [{\cal M}(r)-m]/r,\nonumber 
\end{eqnarray}
where
$$ {\cal M}(r)=r +\frac{q^2}{4r} -\frac{1}{2}\sum^k_{n=2}
        \frac{a_n}{(2n-3)r^{2n-3}}, $$
$m$ and $q$ are the mass and charge of the black hole. The solution 
(\ref{solution1}) is asymptotically flat and has a singularity at $r=0$.
 If $a_n$ are arbitrary, in principle the black hole (\ref{solution1}) 
may have up to $2n-2$ horizons, which are determined by 
the equation $f(r)=0$. When the number of horizons is odd, the 
singularity is spacelike, as that in the Schwarzschild black hole. It 
is timelike as the number of horizons is even, as that of the RN black hole.
Other multihorizon dilaton black holes can also be constructed in
two dimensions.

The multihorizon black holes also appear  in the higher-derivative gravity 
theories. The higher-derivative interactions always  occur in the 
semiclassical quantum gravity and also arise in the effective low-energy
actions of string theories.  Among the higher-derivative gravity theories,
the Lovelock gravity is of a lot of  interest, because it has some attractive
features. For instance, the resulting equations of motion contain no more
than second derivatives of the metric  and it has been proven to be free
of ghosts when expanding about the flat space, evading any problems with
unitarity. The Lovelock Lagrangian is the sum of dimensionally
 extended Euler densities
\begin{equation}
\label{action2}
{\cal L}=\sum^{k}_{n=0}c_n {\cal L}_n,
\end{equation}
where $c_n$ are arbitrary constants and ${\cal L}_n$ is the Euler
density of a $2n$-dimensional manifold
\begin{equation}
{\cal L}_n =2^{-n}\delta^{a_1b_1\cdots a_nb_n}_{c_1d_1\cdots c_nd_n}
           R^{c_1d_1}_{~~~ a_1b_1}\cdots R^{c_nd_n}_{~~~ a_nb_n}.
\end{equation} 
Here the generalized delta function is totally antisymmetric in both sets 
of indices. ${\cal L}_0 =1 $ and hence, $c_0$ is proportional to the 
cosmological constant. ${\cal L}_1$ gives us the Einstein term and
 ${\cal L}_2$ is just the Gauss-Bonnet term. To recover  the result of
general relativity, one usually sets  $c_1>0$. The static, spherically
 symmetric black hole solutions in (\ref{action2}) have been
 found \cite{MS} 
\begin{equation}
\label{solution2}
ds^2=-f(r)~ dt^2 +f^{-1}(r)~dr^2 + r^2 ~ d\Omega^2_{D-2},
\end{equation}
where
$$ f(r)=1-r^2 F(r),$$
and $F(r)$ is determined by solving for the real roots of
the following $k$th-order polynomial equation
$$ P(F)=\sum^{k}_{n=0}\hat c_nF^n =m/r^{D-1}.$$
Here $m$ is an integration constant proportional to the mass 
of the solution, and the coefficients $\hat c_n$ are
$$ \hat c_0=\frac{c_0}{c_1}\frac{1}{(D-1)(D-2)}, 
          \ \ \hat c_1=1, $$
$$ \hat c_n=\frac{c_n}{c_1}\prod ^{2n}_{i=3}(D-i)
            \ \ {\rm for}\ \ n>1.$$
Myers and Simon \cite{MS} analyzed this solution in some detail.
  This solution may be asymptotically flat, de Sitter, 
or anti-de Sitter, depending on the 
parameters in the solution. The maximum number of horizons are computed 
and the nature of singularity at $r=0$ is given in the different classes 
divided according to the parameters (for details see \cite{MS}). 
There they also drawn a schematic Penrose diagram for 
the multihorizon black holes.
Here it should be pointed out that except the singularity at $r=0$,
there may exist other singularities at the finite radii hidden 
by the horizons.

Like the extremal black holes, by adjusting the parameters in solutions 
(\ref{solution1}) and (\ref{solution2}), it is possible 
 that these black hole 
solutions  have degenerate CH's. Here we refer to the inner horizon 
nearest to the event horizon of black holes as the CH. The
 term ``degenerate'' means that the equation $f(r)=0$ has some
 degenerate roots at the CH. Now we discuss  the stability of
 the degenerate CH. For simplicity, we consider the static, spherically 
symmetric black hole  with such a  degenerate CH 
\begin{equation}
\label{solution3}
ds^2= -f(r)dt^2 +f^{-1}(r) dr^2 +r^2d\Omega^2.
\end{equation}
The black hole solution is assumed to may be asymptotically flat, 
anti-de Sitter, or de Sitter. In the asymptotically de Sitter case, the 
cosmological horizon is assumed to exist. The concrete expression
of the metric function $f(r)$, the dimensionality  of  spacetime, and 
the angular part are unimportant to our discussion. Set the 
equation, $f(r)=0$,  has $d (\ge 2)$ degenerate roots at the CH and a 
timelike singularity at $r=0$ is hidden behind the horizons of 
the black hole.

By introducing the null coordinates $v_{\pm}=t \pm r_*$, where 
$r_*=\int f^{-1}(r)dr$,
the black hole spacetime (\ref{solution3}) can be rewritten as
\begin{equation}
\label{metric}
ds^2=-f(r)dv_-dv_+ +r^2d\Omega^2,
\end{equation}
in which $v_+=\infty$ at the CH and the cosmological horizon (if any) and
$v_-=\infty $ at the event horizon of the black hole. 
Using the Kruskal-like coordinates, one can gain the
 regular coordinates at these horizons. To 
investigate the stability of the CH, we use the
 perturbation method \cite{BMM}.
Consider the evolution of a perturbation field denoted by $\Phi$. If the
perturbation field produces a divergent flux at the CH 
measured by a timelike observer, the CH then is unstable due to the
 backreaction of the divergent flux. Otherwise, the CH is stable. 
For  the perturbation field $\Phi$, the resulting flux 
measured by any observer is proportional to the square of the amplitude
\begin{equation}
\label{amp}
{\cal F}=\Phi _{,\alpha}u^{\alpha},
\end{equation}
where $u^{\alpha}$ is the four-velocity of the observer. 
For further discussion on the interaction between the perturbation field 
and the observer see \cite{BO}.
 In the general perturbation analysis it has been found that there is a 
 so-called 
radiation tail at late times \cite{Price}, which is in the leading 
order and blueshifted
at CH's. In the RN black holes, it is just the infinite blueshifted 
radiation tail that causes the CH to be unstable and be converted into
a curvature singularity.   
 
For a radially free-infalling observer falling into the black holes, 
Brady, Moss, and Myers \cite{BMM} have noticed that 
for a regular and reasonable
initial conditions of the perturbations, the observer should see a finite 
flux of radiation at the event horizon of  black holes as well, which 
implies that  
 the perturbation field has the behavior,
 $\Phi _{,v_-}\sim  e^{-\kappa_e v_-}$, near the event horizon, which 
fixes the initial conditions for the outgoing modes. Here
 $\kappa_e$ denotes the surface gravity of the event horizon.  Due to the 
backscattering off the spacetime curvature, the outgoing modes result in 
an additional influx along the CH. The perturbation field then 
has an  additional term \cite{BMM}
\begin{equation}
\label{back}
\Phi  \sim e^{-\kappa_e v_+},
\end{equation}
near the CH. This term is in the subleading order, but it plays a crucial 
role in the instability of the CH in the RN-de Sitter black holes under the 
classical perturbations.

We are now in a position to discuss the stability of the degenerate 
CH in the black hole spacetimes (\ref{solution3}). We first discuss the case
in which the black hole is asymptotically flat.

(i) {\it In the asymptotically flat case}. Both the analytic calculations
 \cite{Price} and the numerical studies of collapse \cite{Gund} give
the perturbation field has a inverse power-law tail at late times. That is
$\Phi$ decays according to
\begin{equation}
\label{tail}
\Phi \sim v_+^{-p},
\end{equation}
where $p$ is a constant and is related to the multipole moment
 $l (\ge s) $ of the perturbation field with spin $s$. Usually it
 has the relation $ p=2l+2 +\gamma$, where $\gamma$ is a constant, 
which is $0$ if there is an initially static perturbation of the
 black holes and $1$ for other cases.
Combining with (\ref{back}), in this case, we have the amplitude of 
the influx measured by the  radially free-infalling observer
\begin{equation}
\label{am1}
{\cal F}_1 \sim v_+^{1+1/d}\left ( v_+^{-p-1} 
      + \beta\  e^{-\kappa_e v_+} \right),
\end{equation}
near the CH, where $\beta $ is a slowly varying function and is a finite 
constant as the CH is approached. The second term
in (\ref{am1}) is finite as $v_+\rightarrow \infty$. The first term is also 
finite if $p\ge 1/d$. In fact the latter alway holds because of $p\ge 2$. 
This finiteness of the influx indicates that the degenerate CH is stable
under the classical perturbations.

(ii) {\it In the asymptotically anti-de Sitter case}. In this case,
 usually one expects that the radiation tail is also inverse 
power-law as in the case of
the asymptotically flat black hole spacetimes. However, recently 
Chan and Mann \cite{Mann}
found that the decay of the perturbation field is complicated and it may be  
 neither the inverse power-law nor exponential  falloff at late times.
 We set the radiation tail having  the behavior $\Phi_{,v_+} \sim g(v_+)$. 
Near the CH, the amplitude then is
\begin{equation}
\label{am2}
{\cal F}_2 \sim v_+^{1+1/d}\left ( g(v_+) 
               + \beta \  e^{-\kappa_e v_+}\right).
\end{equation}
Note that the approximate falloff of the maximal peak 
is weakly exponential and the  the perturbation 
always falls off faster than that of the 
inverse power-law (\ref{tail}) \cite{Mann}. Therefore the influx is also 
finite at the CH for the asymptotically anti-de Sitter black holes, which 
implies that the degenerate CH is also stable against the classical
perturbations.

(iii) {\it In the asymptotically de Sitter case}. In this case, 
 a cosmological horizon with surface gravity $\kappa_c$ is present. When 
 crosses the cosmological horizon, the infalling observer should measure a 
finite flux at the cosmological horizon. It requires that the perturbation
field has the behavior
\begin{equation}
\label{ctail}
\Phi \sim e^{-\kappa_c v_+},
\end{equation}
which fixes the initial conditions of ingoing modes of the perturbations. 
Recent numerical studies of perturbations \cite{BCKL} also give 
the same behavior as (\ref{ctail}). 
 Considering (\ref{back}) and (\ref{ctail}), we have
\begin{equation}
\label{am3}
{\cal F}_3 \sim v_+^{1+1/d}\left (e^{-\kappa_c v_+} +\beta \ e^{-\kappa_e 
     v_+}\right).
\end{equation}
Once again, the influx measured by the infalling observer is finite at the 
CH in this case. Thus, the degenerate CH of black hole spacetimes 
is always stable under the classical perturbations, despite the asymptotic 
behavior of the black holes. When the CH is nondegenerate, 
that is the CH has
a nonvanishing surface gravity $\kappa _i$, the prefactor $v_+^{1+1/d}$
in (\ref{am1}), (\ref{am2}) and (\ref{am3}) is replaced
 by $e^{\kappa_i v_+}$.  The influxes then become  divergent 
 at the CH and hence the nondegenerate CH is classical unstable.

For the classical stable degenerate CH, it is natural to ask whether
it is still stable after taking into account  quantum fluctuations. 
However, to obtain an explicit expression of the vacuum expectation  value 
$\langle T_{\mu\nu}\rangle $ associated with quantum fields is
rather difficult in the physical spacetimes. To see the behavior 
of  quantum fluctuations, it is instructive to discuss the problem in
the  two-dimensional reduction model, $ds^2=-f(r)dv_-dv_+$,
in which the calculation of the expectation 
value $\langle T_{\mu\nu}\rangle $ can be carried out exactly.
  Consider a conformal scalar field propagating in the 
two-dimensional background. Its renormalized expectation value of the 
stress-energy tensor can be expressed as \cite{Davies}
\begin{equation}
\label{tensor}
\langle T_{\mu\nu} \rangle =\Theta _{\mu\nu} + X_{\mu\nu} 
       -(48\pi)^{-1} Rg_{\mu\nu},
\end{equation}
where $X_{\mu\nu} $ depends on the choice of vacuum states, and 
$\Theta _{\mu\nu}$ is
$$ \Theta _{v_+v_+} =(192\pi)^{-1} (2ff''-f'^2),$$
$$\Theta _{v_-v_-}=\Theta _{v_+v_+}, 
\ \ \Theta_{v_-v_+}=\Theta_{v_+v_-}=0.$$ 
Here a prime denotes the derivative with respect to $r$.

To get the  expectation value $\langle T_{\mu\nu}\rangle $ of the quantum
 fluctuations, we have to first specify an appropriate vacuum. 
  When the black hole is  asymptotically flat, there are two  well-known
 vacuum states which are appropriate for our purpose. One is the 
Unruh vacuum state, which can correctly
describe the quantum fields in the spacetimes of black holes formed by 
gravitational collapse. The other is the Israel-Hawking-Hartle vacuum state,
which can describe black holes in thermal 
equilibrium with  the surrounding thermal radiations.
The expectation value $\langle T_{\mu\nu}\rangle $ is finite  at the 
event horizon of black holes for the two vacuum states. However, the energy 
density of the quantum fluctuations measured by the infalling observer has 
the following behavior \cite{Cai} 
\begin{equation}
\label{density1}
\rho _1 \sim v_+^{2+2/d},
\end{equation}
near the CH.  Obviously, this density
 is divergent at the CH, which implies that the degenerate CH is unstable
   under the quantum fluctuations.

When the black hole is asymptotically anti-de Sitter, the
 black hole spacetime is not globally hyperbolic. There is a 
timelike boundary at spatial infinity.
To have a well-posed Cauchy problem, one has to impose 
boundary conditions for quantum fields at spatial infinity $r=\infty$. 
There exist three kinds of appropriately  boundary conditions: Dirichlet,
Neumann,  and  Robin  conditions \cite{Anti}.
 Mode functions satisfying different boundary conditions  
can be defined. The renormalized expectation 
value $\langle T_{\mu\nu}\rangle $ is still given by (\ref{tensor}). 
However, the Unruh vacuum is equivalent to the Israel-Hawking-Hartle one
because of the boundary conditions. Therefore, 
$\langle T_{\mu\nu}\rangle $ is same 
in the two vacuum states. We find that the energy density of the quantum
 fluctuations, measured by the free-infalling observer, is
\begin{equation}
\label{density2}
\rho _2 \sim v_+^{2+2/d},
\end{equation}
near the CH, which has the same behavior as the one in the
 asymptotically flat spacetime (\ref{density1}).

When the black hole spacetime is asymptotically de Sitter, the standard 
vacuum states, Unruh and Israel-Hawking-Hartle states, are both 
inapplicable. An appropriate vacuum is the Markovi\'c-Unruch state 
\cite{Unruh}, in which
the stress-energy tensor is well behaved at the black hole  horizon 
 and the cosmological horizon. In this vacuum state, once again,
we find that the energy density is divergent at the CH as \cite{Mark}
\begin{equation}
\label{density3}
\rho_3 \sim v_+^{2+2/d}.
\end{equation}
Although we have  used  different vacuum states for computing 
the density density of quantum fluctuations in the black hole 
spacetimes possessing different asymptotic behaviors,
 it is easy to see from (\ref{density1}), (\ref{density2}) and
 (\ref{density3})  that the energy densities,
measured by the free-infalling observer, have same divergent behavior 
at the CHs. We therefore expect that the degenerate CH is unstable after 
taking into account the quantum fluctuations, although 
the two-dimensional calculations may have significant differences
 from those in four dimensions.

In summary we have presented one  kind of black holes with degenerate CH's, 
which have vanishing surface gravities there. Despite
the asymptotic behaviors of black holes (flat, anti-de Sitter, or de Sitter),
the degenerate CH is always stable against  the classical perturbations, but 
quantum-mechanically unstable. Thus we give an example of classical stable 
CH in black hole spacetimes. The degenerate CH may occur in some theoretical
models,  as introduced before, however, it remains open whether the 
black holes with the degenerate CH can be formed or not  in the realistic 
circumstances. We hope  that our discussion  is of some significances,
at least in principle.

This work was supported by the KOSEF through the CTP at Seoul National 
University.

\end{document}